# Hamiltonian formulation of fractional kinetics

Sumiyoshi Abe[1,2,3]

[1] Physics Division, College of Information Science and Engineering,
    Huaqiao University, Xiamen 361021, China
[2] Department of Physical Engineering, Mie University, Mie 514-8507, Japan
[3] Institute of Physics, Kazan Federal University, Kazan 420008, Russia

**Abstract.** Fractional kinetic theory plays a vital role in describing anomalous diffusion in terms of complex dynamics generating semi-Markovian processes. Recently, the variational principle and associated Lévy Ansatz have been proposed in order to obtain an analytic solution of the fractional Fokker-Planck equation. Here, based on the action integral introduced in the variational principle, the Hamiltonian formulation is developed for the fractional Fokker-Planck equation. It is shown by the use of Dirac's generalized canonical formalism how the equation can be recast in the Liouville-like form. A specific problem arising from temporal nonlocality of fractional kinetics is nonuniqueness of the Hamiltonian: it has two different forms. The non-equal-time Dirac-bracket relations are set up , and then it is proven that both of the Hamiltonians generate the identical time evolution.

# 1 Introduction

Diffusion is a phenomenon that is ubiquitously observed in variety of systems. The most common may be the one associated with the Gaussian process, which may have its origin in the Einstein-Smoluchowski theory. There, the spatial extension $l$ of a distribution grows in time as $l \sim t^{1/2}$. As well-known, however, there also exist systems, which exhibit

$$l \sim t^{\mu}, \tag{1}$$

where the positive exponent $\mu$ takes the value other than that of normal diffusion $\mu = 1/2$. Such a case is termed anomalous diffusion [1]: subdiffusion if $0 < \mu < 1/2$ and superdiffusion if $\mu > 1/2$. Anomalous diffusion is an important indicator of complexities of systems, as can be expected from the exotic scaling and dimensionality in equation (1). Examples are transport of electrons in amorphous solids [2], motion of microspheres in living cells [3] and of protein chains by MD simulations [4], predator's search of preys [5], pattern of human travel [6] (see also [7]) and volcanic seismicity [8,9], to name a few.

For describing anomalous diffusion, there are at least four major approaches that are frequently discussed in the literature: fractional Brownian motion [10], nonlinear kinetic theory [11], random walks on fractals [12] and fractional kinetic theory [13-16]. Among these, what we study here is fractional kinetic theory.

Fractional kinetics has been attracting much attention in the contexts of continuous time random walks [17] and semi-Markovian processes. Master equations appearing there require the use of fractional calculus, and accordingly it is generically hard to obtain solutions of given fractional kinetic equations in analytic forms. Although numerical analysis is useful in such a situation, approximate analytic solutions are always valuable for getting insights into the physical properties of the systems. Motivated by such a philosophy, we have recently developed the variational principle for fractional kinetic equations [18]. We have shown how this method can explicitly yield useful analytic solutions of fractional Fokker-Planck equations.

The equation discussed in [18] is of the form

$$\frac{\partial p(x,t)}{\partial t} = {}_0D_t^{1-\alpha}\left[-\frac{\partial}{\partial x}(F(x)\,p(x,t)) - \tilde{D}(-\Delta)^{\gamma/2} p(x,t)\right]. \qquad (2)$$

Here, $p(x,t)$ is a normalized probability distribution defined on $(-\infty, \infty) \times [0,T]$, $\tilde{D}$ a generalized diffusion coefficient. The ranges of the fractionality indices of physical interest are

$$0 < \alpha < 1, \qquad 0 < \gamma < 2. \qquad (3)$$

The definitions of the fractional differential operators appearing in equation (2) are as follows [15,19,20]. ${}_0D_t^{1-\alpha}$ denotes the Riemann-Liouville forward fractional differential operator defined by

$$_0D_t^{1-\alpha}[f(t)] = \frac{1}{\Gamma(\alpha)}\frac{d}{dt}\int_0^t ds\,(t-s)^{\alpha-1}f(s) \qquad (4)$$

with $\Gamma(\alpha)$ being the Euler gamma function, which is the adjoint of the backward one

$$_tD_T^{1-\alpha}[f(t)] = \frac{1}{\Gamma(\alpha)}\left(-\frac{d}{dt}\right)\int_t^T ds\,(s-t)^{\alpha-1}f(s) \qquad (5)$$

in the sense that, in the range of $\alpha$ in equation (3), the following equation holds:

$$\int_0^T dt\, f(t)\left\{_0D_t^{1-\alpha}[g(t)]\right\} = \int_0^T dt\,\left\{_tD_T^{1-\alpha}[f(t)]\right\}g(t). \qquad (6)$$

$-(-\Delta)^{\gamma/2}$ is the Riesz fractional Laplacian operator

$$-(-\Delta)^{\gamma/2} = -\frac{1}{2\cos(\pi\gamma/2)}\left[\frac{d^\gamma}{dx^\gamma} + \frac{d^\gamma}{d(-x)^\gamma}\right], \qquad (7)$$

which satisfies $-(-\Delta)^{\gamma/2}\exp(\pm ikx) = -|k|^\gamma \exp(\pm ikx)$ and

$$\int_{-\infty}^\infty dx\,\psi(x)\left\{-(-\Delta)^{\gamma/2}\phi(x)\right\} = \int_{-\infty}^\infty dx\,\left\{-(-\Delta)^{\gamma/2}\psi(x)\right\}\phi(x). \qquad (8)$$

The differential of the drift term in equation (2) could also be fractionalized, in general, but such a generalization does not essentially change the subsequent discussion. The ordinary Fokker-Planck equation corresponds to the limits $\alpha \to 1$ and $\gamma \to 2$, and the familiar Gaussian solution exhibiting normal diffusion is obtained, in particular, in the absence of the drift term. The indices $\alpha$ and $\gamma$ are responsible for the power-law behaviors of the waiting-time and jump distributions of the walker, respectively. Physically, existence of long jumps enhances diffusion, whereas long waiting times suppress diffusion. An approximate variational solution of equation (2) with a periodic drift has been obtained in [18] through the Lévy Ansatz proposed there combined with the Rayleigh-Ritz-like procedure, and the diffusion property has analytically been found to be

$$l \sim t^{\alpha/\gamma}, \tag{9}$$

which is consistent with the physical interpretations of the fractionality indices mentioned above.

From equation (9), we see that the case of $\gamma = 2\alpha$ gives an example of so-called non-Gaussian normal diffusion. We wish to point out that the phenomenon of non-Gaussian normal diffusion itself has been observed in super-cooled liquids [21,22], for example.

In this article, we further advance the discussion in [18] and develop the Hamiltonian formulation of the fractional Fokker-Planck equation (2). We will recast the equation in the form analogous to the Liouville equation

$$\frac{\partial p(x,t)}{\partial t} = \{p(x,t), H(t)\}_D, \qquad (10)$$

where $H(t)$ is the Hamiltonian and the symbol $\{\ ,\ \}_D$ stands for the Dirac bracket [23,24] that generalizes the Poisson bracket. There exists a nontrivial problem originating from temporal nonlocality of the fractional kinetic equation. As can be seen, the Hamiltonian has two different forms. To our knowledge, this point is revealed for the first time through the present work. We will resolve this problem by setting up non-equal-time Dirac-bracket relations and show that both of the Hamiltonians generate the identical time evolution.

This article is organized as follows. In section 2, a succinct review of the variational approach to the fractional Fokker-Planck equation presented in [18, 25] is presented. A problem concerning the normalization condition on a probability distribution is discussed there. In section 3, the Hamiltonian formulation of the fractional Fokker-Planck equation is developed. It is shown how equation (2) can be cast into the form of equation (10). Section 4 is devoted to concluding remarks.

## 2 Variational principle for fractional Fokker-Planck equation

In this section, we present a review of the works in [18,25].

Let $p(x,t)$ be a normalized probability distribution defined on $(-\infty, \infty) \times [0, T]$ and satisfy equation (2). Clearly, equation (2) *as a field equation* does not possess time-reversal invariance, leading to an obstacle for constructing the action functional. A possible way of overcoming this difficulty is to extend the space of the variables by introducing an auxiliary field [26,27], which is denoted here by $\Lambda(x,t)$. Unlike $p(x,t)$, the auxiliary field is not a probability distribution: that is, it need not be normalized and may take negative values. The action reads

$$I[p,\Lambda] = \int_0^T dt \int_{-\infty}^{\infty} dx \, \pounds - \frac{1}{2}\int_{-\infty}^{\infty} dx \left( \Lambda p\big|_{t=0} + \Lambda p\big|_{t=T} \right), \tag{11}$$

where $\pounds$ is the Lagrangian density given by

$$\pounds = \frac{1}{2}\left( \Lambda \frac{\partial p}{\partial t} - \frac{\partial \Lambda}{\partial t} p \right) - \Lambda \, _0D_t^{1-\alpha}\left[ -\frac{\partial}{\partial x}(Fp) - \tilde{D}(-\Delta)^{\gamma/2} p \right]. \tag{12}$$

Henceforth, the range of space integral is omitted for the sake of simplicity.

Since $p(x,t)$ is a probability distribution, it should satisfy the normalization condition $\int dx \, p(x,t) - 1 = 0$ for any $t \in [0,T]$. However, it is actually not necessary to add such a term as a constraint to the action. The reason is as follows. As mentioned, the auxiliary field is not a probability distribution. Its definition admits arbitrariness of a certain kind. Let us consider the following redefinition:

$$\Lambda(x,t) \to \Lambda(x,t) + \int_t^T ds \, \lambda(s), \tag{13}$$

which keeps the final value $\Lambda(x,T)$ unchanged. This arbitrariness may be interpreted as a gauge-theoretical structure. Under the transformation in equation (13), the action changes as

$$I[p,\Lambda] \to I[p,\Lambda] + \int_0^T dt\, \lambda(t) \left\{ \int dx\, p(x,t) - \int dx\, p(x,0) \right\}. \tag{14}$$

On the other hand, the variations with respect to $\Lambda$ and $p$ are

$$\delta_\Lambda I[p,\Lambda] = \int_0^T dt \int dx \left\{ \frac{\partial p}{\partial t} - {_0D_t^{1-\alpha}} \left[ -\frac{\partial}{\partial x}(Fp) - \tilde{D}(-\Delta)^{\gamma/2} p \right] \right\} \delta\Lambda$$

$$- \int dx\, p\,\delta\Lambda \Big|_{t=T}, \tag{15}$$

$$\delta_p I[p,\Lambda] = -\int_0^T dt \int dx \left\{ \frac{\partial \Lambda}{\partial t} + {_tD_T^{1-\alpha}} \left[ F\frac{\partial \Lambda}{\partial x} - \tilde{D}(-\Delta)^{\gamma/2} \Lambda \right] \right\} \delta p$$

$$- \int dx\, \Lambda\,\delta p \Big|_{t=0}, \tag{16}$$

respectively, provided that equations (6) and (8) have been used and integrations by part have been performed under the assumption that the quantity $\Lambda(x,t)F(x)p(x,t)$ vanishes in the limits $x \to \pm\infty$. These equations imply that both $p(x,0)$ and $\Lambda(x,T)$ should be fixed. A fixed initial probability distribution may be $p(x,0) = \delta(x)$, for example. In any case, equation (14) becomes

$$I[p,\Lambda] \to I[p,\Lambda] + \int_0^T dt\, \lambda(t)\left\{\int dx\, p(x,t) - 1\right\}, \tag{17}$$

which shows the $\lambda(t)$ plays a role of the Lagrange multiplier, and the constraint associated with the normalization condition is already implemented in the action.

From the action principle, equations (15) and (16) lead to equation (2) and

$$\frac{\partial \Lambda}{\partial t} = -\,_tD_T^{1-\alpha}\left[F\frac{\partial \Lambda}{\partial x} - \tilde{D}(-\Delta)^{\gamma/2}\Lambda\right], \tag{18}$$

respectively. $\int dx\, \Lambda(x,t)$ is generically not conserved in time, in contrast to the normalization condition on $p(x,t)$. This is why $\Lambda(x,t)$ should not be treated as a probability distribution.

Closing this section, we note apparent nonuniqueness of the Lagrangian density, which is different from the familiar arbitrariness concerning an additional total derivative term and is of importance in the subsequent discussion. A point is that the action principle with another Lagrangian density

$$\pounds^* = \frac{1}{2}\left(\Lambda\frac{\partial p}{\partial t} - \frac{\partial \Lambda}{\partial t}p\right) - \,_tD_T^{1-\alpha}[\Lambda]\left[-\frac{\partial}{\partial x}(Fp) - \tilde{D}(-\Delta)^{\gamma/2}p\right] \tag{19}$$

also gives rise to equations (2) and (18). This is because the action contains time integral and therefore equation (6) can be used for equation (11). On the other hand, the Hamiltonian is given only by space integral of the Hamiltonian density. Therefore, the Hamiltonian has two different forms. This issue will carefully be discussed in the next section.

## 3  Temporal nonlocality, canonical formalism and nonuniqueness of Hamiltonian

In this section, we wish to develop the canonical formalism based on the Lagrangian densities in equations (12) and (19). The canonical momenta conjugate to $p$ and $\Lambda$ are given by $\Pi_p = \partial \pounds / \partial(\partial p / \partial t) = \Lambda/2$ and $\Pi_\Lambda = \partial \pounds / \partial(\partial \Lambda / \partial t) = -p/2$, respectively, which are the same for $\pounds^*$. Since these equations cannot be solved in terms of $\partial p / \partial t$ and $\partial \Lambda / \partial t$, they are regarded as the constraints: $\chi_1 = \Pi_P - \Lambda/2 \approx 0$, $\chi_2 = \Pi_\Lambda + p/2 \approx 0$, where the symbol "$\approx$" denotes weak equality. Appearance of the constraints is simply due to the fact that the Lagrangian densities are of the first order in $\partial p / \partial t$ and $\partial \Lambda / \partial t$. The basic equal-time Poisson bracket relations are given by

$$\{p(x,t), \Pi_p(x',t)\} = \delta(x-x'), \tag{20}$$

$$\{\Lambda(x,t), \Pi_\Lambda(x',t)\} = \delta(x-x'). \tag{21}$$

The constraints are then seen to be of the second class [23,24], since $\{\chi_1(x,t), \chi_2(x',t)\} = -\delta(x-x')$, which does not weakly vanish. Then, it is standard to eliminate the second-class constraints by introducing the equal-time Dirac bracket given

by

$$\{A(x,t), B(x',t)\}_D = \{A(x,t), B(x',t)\} - \sum_{i,j=1}^{2} \iint dy\,dy' \{A(x,t), \chi_i(y,t)\}$$
$$\times C_{ij}(y,y')\{\chi_j(y',t), B(x',t)\}, \quad (22)$$

where $C_{ij}(y,y')$ is the quantity satisfying $\sum_{k=1}^{2} \int dy \{\chi_i(x,t), \chi_k(y,t)\} C_{kj}(y,y')$ $= \delta_{ij}\delta(x-y')$: that is, $C_{11}(y,y') = C_{22}(y,y') = 0$, $C_{12}(y,y') = -C_{21}(y,y') = \delta(y-y')$. From equations (20) and (21), holds $\{\chi_i(x,t), \chi_j(x',t)\}_D = 0$ ($i,j=1,2$), and so the second-class constraints can be regarded as the strong equations (i.e., the identities). The basic equal-time Dirac-bracket relations are found to be

$$\{p(x,t), p(x',t)\}_D = 0, \tag{23}$$

$$\{\Lambda(x,t), \Lambda(x',t)\}_D = 0, \tag{24}$$

$$\{p(x,t), \Lambda(x',t)\}_D = \delta(x-x'). \tag{25}$$

Equation (25) implies that $p$ and $\Lambda$ are canonically conjugate to each other with respect to the Dirac bracket.

Our purpose is to represent equations (2) and (18) in the Liouville-like form: $\partial A(x,t)/\partial t = \{A(x,t), H(t)\}_D$. The Hamiltonian $H(t)$ is space integral of the Hamiltonian density given by the Legendre transformation: (Hamiltonian density) $= \Pi_p \, \partial p / \partial t + \Pi_\Lambda \, \partial \Lambda / \partial t -$ (Lagrangian density). However, as mentioned in the preceding section, there exist two different forms for the Lagrangian density as in equations (12) and (19), and accordingly, the Hamiltonian also has two different forms:

$$H(t) = \int dx \, \Lambda \,_0D_t^{1-\alpha}[L p], \qquad (26)$$

$$H^*(t) = \int dx \, \left(_tD_T^{1-\alpha}[L^\dagger \Lambda]\right) p, \qquad (27)$$

respectively, where

$$L \equiv -\frac{\partial}{\partial x} F(x) - \tilde{D}\,(-\Delta)^{\gamma/2}, \qquad (28)$$

$$L^\dagger \equiv F(x)\frac{\partial}{\partial x} - \tilde{D}\,(-\Delta)^{\gamma/2}, \qquad (29)$$

which are adjoint to each other under the inner product: $\int dx \, \psi(x)(L\phi(x)) = \int dx \, (L^\dagger \psi(x))\phi(x)$, provided that $\psi(x) F(x) \phi(x)$ should vanish in the limits $x \to \pm\infty$. From equations (23)-(25), we immediately have

$$\frac{\partial p(x,t)}{\partial t} = \{p(x,t), H(t)\}_D$$
$$= {}_0D_t^{1-\alpha}[L p(x,t)], \qquad (30)$$

$$\frac{\partial \Lambda(x,t)}{\partial t} = \left\{ \Lambda(x,t), H^*(t) \right\}_D$$

$$= -{}_t D_T^{1-\alpha} \left[ L^\dagger \Lambda(x,t) \right], \qquad (31)$$

which are in fact equations (2) and (18), respectively. We note that, in the above, we have imposed the following conditions: $\left\{ p(x,t), {}_0 D_t^{1-\alpha}[p(x',t)] \right\}_D = 0$, $\left\{ \Lambda(x,t), {}_t D_T^{1-\alpha}[\Lambda(x',t)] \right\}_D = 0$, which are fulfilled if the non-equal-time Dirac-bracket relations

$$\left\{ p(x,t), p(x',\tau) \right\}_D = 0, \qquad (32)$$

$$\left\{ \Lambda(x,t), \Lambda(x',\tau) \right\}_D = 0, \qquad (33)$$

hold.

To ascertain that the two different forms of the Hamiltonian in equations (26) and (27) generate identical time evolution, it is necessary to show $\left\{ p(x,t), H^*(t) \right\}_D$ and $\left\{ \Lambda(x,t), H(t) \right\}_D$ also give rise to equations (30) and (31), respectively. For this purpose, it is necessary to calculate $\left\{ p(x,t), \Lambda(x',\tau) \right\}_D$. This is a specific point originating from temporal nonlocality of fractional kinetics. Since the Dirac-bracket relations should be consistent with the *equations of motion*, equations (2) and (18) can be used in such a calculation. Let us integrate the both sides in equation (2) with respect to time:

$$p(x,t) = p(x,0) + \frac{1}{\Gamma(\alpha)} \int_0^t d\tau \, (t-\tau)^{\alpha-1} L p(x,\tau). \tag{34}$$

This equation is formally solved by the use of the iteration method, yielding

$$p(x,t) = E_\alpha(t^\alpha L) p(x,0), \tag{35}$$

where $E_\alpha(z)$ is the Mittag-Leffler function $E_\alpha(z) = \sum_{n=0}^{\infty} z^n / \Gamma(n\alpha + 1)$, which is a special case of the generalized Mittag-Leffler function $E_{\alpha,\beta}(z) = \sum_{n=0}^{\infty} z^n / \Gamma(n\alpha + \beta)$: that is, $E_\alpha(z) = E_{\alpha,1}(z)$. From equation (35), it follows that

$$p(x,t) = E_\alpha(t^\alpha L) E_\alpha^{-1}(\tau^\alpha L) p(x,\tau), \tag{36}$$

which generalizes equation (25) to

$$\{p(x,t), \Lambda(x',\tau)\}_D = E_\alpha(t^\alpha L) E_\alpha^{-1}(\tau^\alpha L) \delta(x-x'), \tag{37}$$

where $E_\alpha^{-1}(\tau^\alpha L)$ is the inverse operator of $E_\alpha(\tau^\alpha L)$. It can be found that the following relation holds for the operator $E_\alpha(t^\alpha L)$:

$$\frac{\partial}{\partial t} E_\alpha(t^\alpha L) = {}_0D_t^{1-\alpha}[E_\alpha(t^\alpha L)] L \equiv t^{\alpha-1} L E_{\alpha,\alpha}(t^\alpha L). \tag{38}$$

Equations (32), (33) and (37) stipulate the basic non-equal-time Dirac-bracket relations.

Now, $\{p(x,t), H^*(t)\}_D$ is calculated as follows:

$$\{p(x,t), H^*(t)\}_D = \int dx' \, p(x',t) \{p(x,t), {}_tD_T^{1-\alpha}[L'^\dagger \Lambda(x',t)]\}_D$$

$$= -\int dx' \, p(x',t) \frac{\partial}{\partial \tau} \{p(x,t), \Lambda(x',\tau)\}_D \Big|_{\tau=t}$$

$$= -\int dx' \, p(x',t) E_\alpha(t^\alpha L) \frac{\partial}{\partial \tau} E_\alpha^{-1}(\tau^\alpha L) \Big|_{\tau=t} \delta(x-x')$$

$$= \frac{\partial E_\alpha(t^\alpha L)}{\partial t} p(x,0)$$

$$= {}_0D_t^{1-\alpha}[L p(x,t)] \,, \tag{39}$$

where equations (18), (35) and (38) have been employed. Thus, we see that equation (30) can also be written as

$$\frac{\partial p(x,t)}{\partial t} = \{p(x,t), H^*(t)\}_D. \tag{40}$$

In a similar way, equation (31) can also be shown to coincide with

$$\frac{\partial \Lambda(x,t)}{\partial t} = \{\Lambda(x,t), H(t)\}_D. \tag{41}$$

Therefore, we conclude that both $H(t)$ and $H^*(t)$, in fact, generate the identical time evolution.

## 4 Concluding remarks

We have developed the Hamiltonian formulation of the fractional Fokker-Planck equation and have recast the equation in the Liouville-like form. The problems arising from temporal nonlocality and nonuniqueness of the Hamiltonian have been resolved.

In the present formulation, no strict restrictions are put on the auxiliary field $\Lambda(x,t)$. This freedom is advantageous for discussions about symmetry. Since it is canonically conjugate to $p(x,t)$ with respect to the Dirac bracket, it can be used, for example, for construction of the generator of dilatation transformation. Symmetry (or approximate symmetry) associated with such a transformation is directly related to the scale invariance of the system [28].

**Acknowledgment**

This work has been supported in part by a grant from National Natural Science Foundation of China (No. 11775084) and Grant-in-Aid for Scientific Research from the Japan Society for the Promotion of Science (No. 26400391 and No. 16K05484), and by the Program of Competitive Growth of Kazan Federal University from the Ministry of Education and Science of the Russian Federation.